\documentclass[a4paper, 11pt]{article}

\usepackage{a4wide}
\usepackage{amsmath, amsthm}
\usepackage{amsfonts}
\usepackage{amssymb}
\usepackage{enumerate,enumitem}
\usepackage[bookmarks=false,colorlinks=true,linkcolor=black,citecolor=black,dvips=true]{hyperref}
\usepackage[latin1]{inputenc}
\usepackage[english]{babel}
\usepackage{graphicx}      

\graphicspath{{fig-adis/}}

\newcommand{\cf}{{\mathcal F}}
\newcommand{\cn}{{\mathcal N}}
\newcommand{\E}{{\mathbb E}}
\renewcommand{\L}{{\mathbb L}}

\renewcommand{\P}{{\mathbb P}}
\newcommand{\R}{{\mathbb R}}

\newcommand{\s}{\star}
\renewcommand{\t}{\theta}
\newcommand{\ts}{\theta^\star}

\newcommand{\ind}[1]{{\bf 1}_{\{#1\}}}
\newcommand{\abs}[1]{|#1|}
\newcommand{\norm}[1]{\|#1\|}
\newcommand{\Var}[0]{\mathop{\rm Var}\nolimits}
\newcommand{\expp}[1]{\mathop {\mathrm{e}^{ #1}}}

\emergencystretch=\hsize
\makeatletter
\newcounter{hypo}
\newcommand*{\dohypo}{\textbf{(A\thehypo)}}
\newenvironment{hypo}[1][]{%
  
  \refstepcounter{hypo}
  \list{}{%
    \settowidth{\labelwidth}{\dohypo}%
    \setlength{\labelsep}{10pt}%
    \setlength{\leftmargin}{\labelwidth}
    \advance\leftmargin\labelsep%
  }%
\item[\dohypo  #1]%
}{%
  \endlist
}
\newcommand*{\newhypo}{
  \refstepcounter{hypo}
\item[\dohypo]}

\def\hypref#1{\hyperref[#1]{(A\ref*{#1})}}
\def\hypreff#1#2{\hyperref[#2]{(A\ref*{#1}-{\it \ref*{#2})}}}
\makeatother

\newtheorem{theorem}{Theorem}[section]
\newtheorem{algo}[theorem]{Algorithm}

\newtheorem{proposition}[theorem]{Proposition}
\newtheorem{corollary}[theorem]{Corollary}
\newtheorem{remark}[theorem]{Remark}

\begin{document}

\title{A framework for adaptive Monte-Carlo procedures}

\date{\today}
\author{Bernard Lapeyre\footnote{
    Université Paris-Est, CERMICS, Projet MathFi
    ENPC-INRIA-UMLV, 6 et 8 avenue Blaise Pascal, 77455 Marne La Vallée, Cedex
    2, France , e-mail : bernard.lapeyre@enpc.fr.} \and Jérôme Lelong\footnote{
    Laboratoire Jean Kuntzmann, Universit\'e de Grenoble et CNRS, BP 53, 
    38041 Grenoble C\'edex 9, France, 
    e-mail : jerome.lelong@imag.fr}}

\maketitle

\begin{abstract} Adaptive Monte Carlo methods are recent variance reduction
  techniques. In this work, we propose a mathematical setting which greatly
  relaxes the assumptions needed by for the adaptive importance sampling
  techniques presented in
  \cite{vazquez-abad98,Su02optimalimportance,MR2054568,arouna03:_robbin_monro}.
  We establish the convergence and asymptotic normality of the adaptive Monte
  Carlo estimator under local assumptions which are easily verifiable in
  practice. We present one way of approximating the optimal importance
  sampling parameter using a randomly truncated stochastic algorithm. Finally,
  we apply this technique to some examples of valuation of financial derivatives.
\end{abstract}

\section{Introduction}

Monte-Carlo methods aim at computing the expectation $\E(Z)$ of a real-valued
random variable $Z$ using samples along the law of $Z$. In this work, we focus
on cases where there exists a parametric representation of the expectation
\begin{equation}
  \label{parrep}
  \E(Z) = \E\left(H(\t,X)\right) \quad \mbox{for all } \t \in \R^d,
\end{equation}
where $X$ is a random vector with values in $\R^m$ and $H : \R^d \times \R^m
\longmapsto \R$ is a measurable function satisfying $\E\abs{H(\t, X)} < \infty$
for all $\t \in \R^d$. We also impose that 
\begin{equation}
  \label{parrepL2}
  \t \longmapsto v(\t) = \Var(H(\t,X)) \mbox{ is finite for all }
  \t \in \R^d,
\end{equation}
We want to make the most of this free parameter $\t$ to settle an automatic variance
reduction method, see~\cite{jourdain09} for a recent survey on adaptive variance
reduction. It consists in first finding a minimiser $\t^\s$ of the variance $v$ and
then in plugging it into a Monte Carlo method with a narrower confidence interval.
This technique heavily relies on the ability to find a parametric representation and
to effectively minimise the function $v$. Many papers have been written on how to
construct parametric representations $H(\t, X)$ for several kinds of random variables
$Z$. We mainly have in mind examples based on control variates (see
\cite{MR1999614,kimhenderson,kim04:_adapt_contr_variat}) or importance sampling (see
\cite{vazquez-abad98,Su02optimalimportance,MR2054568,arouna03:_robbin_monro}).  We
refer the reader to section~\ref{sec-ex} for a presentation of a few examples.

Assume we have a parametric representation of the form $H(\t, X)$ satisfying
Equations~\eqref{parrep} and~\eqref{parrepL2}. Let $(X_n)_n$ be an independent and
identically distributed sequence of random vectors following the law of $X$. Assume
we know how to use the sequence $(X_n)_n$ to build an estimator $\t_n$ of $\t^\s$
adapted to the filtration $\cf_n = \sigma(X_1, \dots, X_n)$. Once such an
approximation is available, there are at least two ways of using it to devise a
variance reduction method.

\paragraph{The non-adaptive algorithm}

\begin{algo}[Non adaptive importance sampling (NADIS)]
  \label{algo:not-coupled}
  Let $n$ be the number of samples used for the Monte Carlo computation.  Draw
  a second set of $n$ samples $(X'_1, \dots, X'_n)$ independent of $(X_1,
  \dots, X_n)$ and compute
    \begin{equation*}
      \bar \xi_n = \frac{1}{n} \sum_{i=1}^n H(\t_n, X'_i).
    \end{equation*}
\end{algo}
Since the sequence $(\t_n)_n$ converges to $\t^*$, the convergence
of $(\bar \xi_n)_n$ to $\E(Z)$ ensues from the strong law of large numbers and
the sequence $(\bar \xi_n)_n$ satisfies a central limit theorem
\[
\sqrt{n} (\bar \xi_n - \E(Z)) \xrightarrow[n \rightarrow \infty]{law} \cn\left(0,
  v(\t^\s) \right).
\]
This algorithm has been studied in
\cite{Su02optimalimportance,arouna03:_robbin_monro} and required $2n$
samples. It may use less than $2n$ samples if the estimation of $\ts$ is
performed on a smaller number of samples but then it raises the question of how
many samples to use. 

\paragraph{The adaptive algorithm}

The adaptive approach is to use the {\it same samples $(X_1, \dots, X_n)$} to
compute $\t_n$ and the Monte Carlo estimator. Compared to the sequential
algorithm, the adaptive one uses half of the samples.
\begin{algo}[Adaptive Importance Sampling (ADIS)]
  \label{algo:coupled}
  Let $n$ be the number of samples used for the Monte Carlo computation. \\
  For $\t_0$ fixed in $\R^d$, compute
  \begin{equation}
    \label{adaptive-estim}
    \xi_n = \frac{1}{n} \sum_{i=1}^n H(\t_{i-1}, X_i).
  \end{equation}
\end{algo}
Note that the sequence $(\xi_i)_i$ can be written in a recursive manner so that
it can be updated online each time a new iterate $\t_i$ is drawn
\begin{equation*}
   \xi_{i+1} = \frac{i}{i+1} \xi_i + \frac{1}{i+1} H(\t_i, X_{i+1}),
  \quad \mbox{with $\xi_0=0$}.
\end{equation*}
Being able to update the sequence $(\xi_i)_i$ online has the advantage that
there is no need to store the whole sequence $(X_1,\dots,X_n)$ for
computing $\xi_n$.  This adaptive algorithm was first studied in
\cite{MR2054568} in which the author studied the convergence of the
sequence $(\xi_n)_n$ under assumptions to be verified along the path
$(\t_n)_n$ which makes them hard to check in practise. In this article, we
prove a new convergence result under local integrability conditions on the
function $H$, namely we impose that for any compact subset $K$ of $\R^n$,
$\sup_{\t \in K} \E(|H(\t, X)|^2) < \infty$.  We refer the reader to
section~\ref{sec-conv} for a precise statement and proof of these results.
We want to emphasize that such assumptions only involving properties of the
function $H$ and not of the sequence $(\t_n)_n$ are far easier to check in
practice.

Sofar, we have assumed that we knew how to devise a convergent estimator of $\t^\s$,
but this may not be so simple as when no closed form expression is available for
$\E\left(H(\t,X)\right)$, there is hardly no chance that the function $v$ can be
computed explicitly. Henceforth, it is needed to approximate $\t^\s$ without being
able to compute the variance itself.  In this work, we recall the methodology based
on stochastic approximation developed
in~\cite{Su02optimalimportance,vazquez-abad98,arouna03:_robbin_monro} to estimate
$\t^\s$ using some stochastic gradient style algorithms.  We aim at applying this
methodology to the evaluation of financial derivatives and the main difficulty in
approximating $\t^\s$ comes from the non-boundedness of the payoff functions usually
considered and consequently the non-boundedness of the $H$ functions. To encompass
this problem, several authors as in~\cite{vazquez-abad98,Su02optimalimportance}
restrict the parameter $\t$ to lie in a compact set, which is obviously unknown in
practice; therefore, this compact set will have to be quite large. Although, it
permits to prove the theoretical convergence of the Robbins-Monro algorithm it does
not help to build a numerically convergent estimator of $\t^\s$. We all know that the
true convergence of stochastic algorithms highly relies on the fine tuning of the
gain sequence which reveals to be very difficult when dealing with an artificially
bounded parameter set. 

In this work following~\cite{arouna03:_robbin_monro}, we would rather use a randomly
truncated algorithm which is known to converge for a much wider class of functions.
We give a unified framework with easily verifiable assumptions under which
Algorithm~\ref{algo:coupled} converges and satisfies a central limit theorem. Then,
we combine this convergence result with the new results on randomly truncated
stochastic algorithm from~\cite{lelong_as} to revisit the adaptive algorithm in the
Gaussian framework studied in~\cite{MR2054568}.

The paper is organised as follows.  In Section~\ref{sec-maths}, we focus on the
mathematical foundation of the method and give both a strong law of large numbers and
a central limit theorem for the adaptive estimator under weak assumptions.  In
Section~\ref{sec-sa}, we present  one way of constructing a convergent estimator of
$\t^\s$ and recall some recent results on stochastic approximation.  Then, we give in
Section~\ref{sec-ex} some examples of how to construct a parametric estimator using
importance sampling or other more elaborate transformations.  Finally, we illustrate
the convergence results obtained in Section~\ref{sec-maths} on numerical examples
coming from financial problems.

\section{Mathematical foundations of the method}
\label{sec-maths}

\textsc{Notations}:~ \begin{itemize}
  \item We encode any elements of $\R^m$ as column vectors.
  \item If $x \in \R^m$, $x^*$ is a row vector. We use the ``${}^*$''
    notation to denote the transpose operator for vectors and matrices.
  \item If $x, y \in \R^m$, $x \cdot y$ denotes the Euclidean scalar
    product of $x$ and $y$ and the associated norm is denoted by $|\cdot|$.
\end{itemize}

In this section, $(X_n)_{n \ge 1}$ is an i.i.d. sequence following the law of
$X$ and we introduce the $\sigma-$algebra $\cf_n$ it generates $\cf_n =
\sigma(X_1, \dots, X_n)$. For technical reasons, we assume that the variance
$v$ does not vanish, i.e. $\inf_{\t \in \R^n} v(\t) > 0$. If such is not the
case, it means that we are actually in a better situation as far as variance
reduction is concerned but it does not fit in our framework.

\subsection{An adaptive strong law of large numbers}
\label{sec-conv}

\begin{theorem}[Adaptive strong law of large numbers] \label{adapslln} Assume
  Equation~\eqref{parrep} and~\eqref{parrepL2} hold.  Let $(\t_n)_{n\geq 0}$
  be a $(\cf_n)-$adapted sequence with values in $\R^d$ such that for all 
  $n \ge 0$, $\t_n < \infty$ a.s and for any compact subset $K \subset \R^d$,
  $\sup_{\t \in K}  \E(\abs{H (\t,X)}^2) < \infty$. If
  \begin{equation}
    \label{vbounds}
    \inf_{\t\in\R^d} v(\t) > 0 \qquad \mbox{and} \qquad
    \frac{1}{n} \sum_{k=0}^n v(\t_{k}) < \infty \quad \mbox{a.s.}, 
  \end{equation}
  then  $\xi_n$ converges a.s. to $\E(Z)$.
\end{theorem}
\begin{proof} For any $p \ge 0$, we define $\tau_p = \inf\{ k \ge 0;
  \abs{\t_k} \ge p\}$. The sequence $(\tau_p)_p$ is an increasing sequence of
  $(\cf_n)-$stopping times such that $\lim_{p \rightarrow \infty} \tau_p
  \uparrow \infty$ a.s.. Let $M_n = \sum_{i=0}^{n-1} H (\t_i,X_{i+1}) -
  \E(Z)$. We introduce $M^{\tau_p}_n = M_{\tau_p \wedge n}$ defined by
  \begin{equation*}
    M^{\tau_p}_n = \sum_{i=0}^{n-1 \wedge \tau_p} H (\t_i,X_{i+1}) - \E(Z) =
    \sum_{i=0}^{n-1} (H (\t_i,X_{i+1}) - \E(Z) )\ind{i \le \tau_p}.
  \end{equation*}
  $\E(\abs{H (\t_i,X_{i+1})- \E(Z)}^2 \ind{i \le \tau_p}) \le \E(\ind{i \le
    \tau_p} \E(\abs{H (\t,X) - \E(Z)}^2)_{\t = \t_i})$. On the set $\{i \le
  \tau_p\}$, the conditional expectation is bounded from above by
  $\sup_{\abs{\t} \le p} v(\t)$.  Hence, the sequence $(M^{\tau_p}_n)_n$ is
  square integrable and it is obvious that $(M^{\tau_p}_n)_n$ is a martingale,
  which means that the sequence $(M_n)_n$ is a locally square integrable
  martingale (i.e. a local martingale which is locally square integrable).
  \begin{equation*}
    \langle M \rangle_n = \sum_{i=0}^{n-1} \E((H
    (\t_i,X_{i+1}) -\E(Z))^2 | \cf_i) = \sum_{i=0}^{n-1} v(\t_i).
  \end{equation*}
  By Condition~\eqref{vbounds}, we have a.s.  $\lim\sup_n \frac{1}{n} \langle
  M \rangle_n < \infty$ and $\lim\inf_n \frac{1}{n} \langle
  M \rangle_n > 0$. Applying the strong law of large numbers for
  locally $\L^2$ martingales (see \cite{lepingle78_conv_mart_loc}) yields the
  result.
\end{proof}

The sequence $(\t_n)_n$ can be any sequence adapted to $(X_n)_{n \ge 1}$
convergent or not. For instance, $(\t_n)_n$ can be an ergodic Markov chain
distributed around the minimizer $\t^\s$ such as Monte Carlo Markov Chain
algorithms.

\begin{remark}
  When the sequence $(\t_n)_{n\geq 0}$ converges a.s. to a deterministic
  constant $\t_\infty$, it is sufficient to assume that $v$ is continuous at
  $\t_\infty$ and $v(\t_\infty) > 0$ to ensure that Condition~\eqref{vbounds}
  is satisfied. Note that there is no need to impose that $\t_\infty = \t^\s$
  although it is undoubtedly wished in practice.  For instance, $\t_\infty$
  can be an approximation of $\t^\s$ obtained either by heuristic arguments
  such as large deviations.
\end{remark}

\subsection{A Central limit theorem for the adaptive strong law of large
  numbers}

To derive a central limit theorem for the adaptive estimator $\xi_n$, we need
a central limit theorem for locally square integrable martingales, whose
convergence rate has been extensively studied. We refer to the works of
Rebolledo~\cite{rebolledo80}, Jacod and Shiryaev~\cite{jacodshiryaev87}, Hall
and Heyde~\cite{MR624435} and Whitt~\cite{whitt07} to find different
statements of central limit theorems for locally square integrable càdlàg
martingales in continuous time, from which theorems can easily be deduced for
discrete time locally square integrable martingales.

\begin{theorem}
  \label{adaptcl} Assume Equation~\eqref{parrep} and~\eqref{parrepL2} hold.
  Let $(\t_n)_{n\geq 0}$ be a $\cf_n-$adapted sequence with values in $\R^d$
  such that for all $n \ge 0$, $\t_n < \infty$ a.s and converging to some
  deterministic value $\t_\infty$. Assume there exists $\eta > 0$ such that
  the function $s_{2+\eta} : \t\in \R^d \longmapsto
  \E\left(\abs{H(\t,X)}^{2+\eta}\right)$ is finite for all $\t \in \R^d$ and
  continuous at $\t_\infty$. Moreover, if $v$ is continuous at $\t_\infty$ and
  $v(\t_\infty) >0$, then, $\sqrt{n} (\xi_n - \E(Z) ) \xrightarrow[]{law} \cn
  (0,v(\t_\infty))$.
\end{theorem}
\begin{proof}
  We know from the proof of Theorem~\ref{adapslln} that $M_n =
  \sum_{i=0}^{n-1}  H (\t_i,X_{i+1}) - \E(Z)$ is a locally square integrable
  martingale and that $\frac{1}{n} \langle M \rangle_n$ converges
  a.s. to $v(\t_\infty)$. 
  \begin{equation*} \frac{1}{n} \sum_{i=0}^{n-1} \E( |H (\t_i,X_{i+1}) -
    \E(Z)|^{2+\eta} | \cf_i) \le c \left(\frac{1}{n} \sum_{i=0}^{n-1}
      s_{2+\eta} (\t_i) + \E(Z)^{2+\eta}\right).
  \end{equation*}
  The term on the r.h.s is bounded thanks to the continuity of $s_{2+\eta}$ at
  $\t_\infty$. Hence, the local martingale $(M_n)_n$ satisfies Lindeberg's
  condition. The result ensues from the central limit theorem for locally
  $\L^2$ martingales.
\end{proof}

\begin{corollary}[Effective central limit theorem with confidence interval]
  \label{varestim} Assume Equation~\eqref{parrep} and~\eqref{parrepL2} hold.
  Let $(\t_n)_{n\geq 0}$ be a $\cf_n-$adapted sequence with values in $\R^d$
  such that for all $n \ge 0$, $\t_n < \infty$ a.s and converging to some
  deterministic value $\t_\infty$. Assume there exists $\eta > 0$ such that
  the function $s_{4+\eta} : \t\in \R^d \longmapsto
  \E\left(\abs{H(\t,X)}^{4+\eta}\right)$ is finite for all $\t \in \R^d$ and
  continuous at $\t_\infty$.  Then, $\sigma_n^2 = \frac{1}{n} \sum_{i=0}^{n-1}
  H (\t_i,X_{i+1})^2 - \xi_n^2 \xrightarrow[]{a.s.} v(\t_\infty)$.  If moreover
  $v(\t_\infty) >0$, then $\frac{\sqrt{n}}{\sigma_n} (\xi_n - \E(Z) )
  \xrightarrow[n \rightarrow +\infty]{law} \cn (0,1)$.
\end{corollary}
\begin{remark} Even if $v(\t_\infty) >0$, $\sigma_n$ may take negative values
  for $n$ small. This corollary is really essential from a practical point of
  view because it proves that confidence intervals can be built as in the case
  of a crude Monte Carlo procedure. The only difference lies in the way of
  approximating the asymptotic variance.
\end{remark}

The assumptions of Theorem~\ref{adaptcl} are fairly easy to check in practice
since they are formulated independently of the sequence $(\t_n)_n$. When
$\t_\infty = \t^\s$, which is nonetheless not required, the limiting variance
is optimal in the sense that a crude Monte Carlo computation with the optimal
parameter $\t^\s$ would have lead to the same limiting variance. These
assumptions are satisfied in the frameworks introduced in
Section~\ref{sec-ex}.

\section{Estimation of the optimal variance parameter}
\label{sec-sa}

From Theorem~\ref{adapslln} and Theorem~\ref{adaptcl}, we know that if
we can construct a convergent estimator $(\t_n)_n$ of $\t^\s$, the adaptive
estimator $\xi_n$ is a convergent and asymptotically normal estimator of the
expectation $\E(Z)$. The challenging issue is now to propose an automatic way
of approximating the minimiser $\t^\s$ of $v(\t) = \E(H(\t, X)^2) - \E(Z)^2$. In the
following, we will assume that $v$ is strictly convex, goes to infinity at
infinity and is continuously differentiable. Moreover, we assume that $\nabla
v$ admits a representation as an expectation
\begin{equation*}
  \nabla v(\t) = \E(U(\t, X)),
\end{equation*}
where $U : \R^d \times \R^m \longmapsto \R^d$ is a measurable and integrable
function. We could see in the examples developed in Section~\ref{sec-ex} that
these conditions are very easily satisfied. Stochastic algorithms such as the
Robbins Monro algorithm~(see~\cite{MR0042668}) are perfectly well suited to
estimate quantities defined as the root of an expectation. Because for the
applications we are targeting we cannot impose that $\E(|U(\t, X)|^2) < c
(1+|\t|^2)$, the Robbins-Monro algorithm will fail to converge and we need a
more robust algorithm.  This will naturally lead us to consider randomly
truncated stochastic algorithms as introduced by Chen et al.
\cite{chen86:_stoch_approx_proced}. When dealing with stochastic
approximations, the idea of averaging the iterates comes out quite
naturally to smooth the trajectories, see Section~\ref{sec:average}.

\subsection{Randomly truncated stochastic algorithms}
\label{sec-chen}

Let ${(X_n)}_{n \geq 1}$ be an i.i.d sequence of random variables following
the law of $X$ and ${(\gamma_{n})}_{n \geq 1}$ be a decreasing sequence of a
positive real numbers satisfying
\begin{equation}
  \label{gain}
  \sum_n \gamma_n = \infty \quad \mbox{and} \quad \sum_n \gamma_n^2 < \infty.
\end{equation}
The sequence $(\gamma_n)_n$ is often called the gain sequence or the step
sequence.  We define the $\sigma-$field $\cf_n = \sigma(X_k, \; k \leq n)$. We
introduce an increasing sequence of compact sets $(K_j)_j$ of $\R^d$
\begin{equation}
  \label{eq:chen-compact-def}
  \bigcup_{n=0}^{\infty} K_n \: = \: \R^d \quad \mbox{and} \quad K_{n}
  \subsetneq \mathring{K}_{n+1}
\end{equation}
Now, we can present the randomly truncated stochastic algorithm introduced
in~\cite{chen86:_stoch_approx_proced}, which essentially consists in a
truncation of the Robbins Monro algorithm on an increasing sequence of compact
sets.  For $\t_0 \in K_0$ and $\alpha_0 = 0$, we define the sequences of
random variables ${(\t_n)}_n$ and ${(\alpha_n)}_n$ by
\begin{equation}
  \label{eq:chen}
  \begin{cases}
    & \t_{n + \frac{1}{2}}  = \t_{n} - \gamma_{n+1} U(\t_n, X_{n+1}),\\
    \text{if $\t_{n + \frac{1}{2}} \in \mathcal{K}_{\alpha_n}$} & \t_{n+1} =
    \t_{n + \frac{1}{2}} \quad \mbox{ and } \quad \alpha_{n+1} = \alpha_n, \\
    \text{if $\t_{n + \frac{1}{2}} \notin \mathcal{K}_{\alpha_n}$} & \t_{n+1}
    = \t_{0}  \quad \mbox{ and } \quad \alpha_{n+1} = \alpha_n + 1. 
  \end{cases}
\end{equation}
$\t_{n + \frac{1}{2}}$ is the new sample we draw, either we accept it and set
$\t_{n+1}=\t_{n+\frac{1}{2}}$ or we reject it and reset the algorithm to
$\t_{0}$ when it tries to jump too far ahead in a single step. Note that
$\t_{n + \frac{1}{2}}$ is actually drawn along the dynamics of the Robbins
Monro algorithm and either we accept it as the new iterate or we reject it
when the algorithm tries to jump to far ahead and in this case we reset the
new iterate to $\t_0$. In the following, we write Equation~\eqref{eq:chen} in
a more condensed form
\begin{equation}
  \label{chen2}
  \t_{n + 1} = {\cal T}_{K_{\alpha_n}} \left(\t_{n} - \gamma_{n+1} U(\t_n, X_{n+1})\right),
\end{equation}
where ${\cal T}_{K_{\alpha_n}}$ denotes the truncation on the compact sets
$K_{\alpha_n}$.

The use of truncations enables to relax the hypotheses required to ensure the
convergence. From the recent results of~\cite{lelong_as}, we can state the
following convergence result
\begin{theorem}
  \label{thm-chen}
  Assume that 
  \begin{hypo}
    \label{unique-rm} $\nabla v$ is continuous and there exists a unique
    $\t^\s$ s.t. $\nabla v(\t^\s) = 0$ and $\forall \; \t \neq \t^\s$,
    $(\nabla v(\t) \,|\, \t - \t^\s) >0$.
    \newhypo\label{localbound}
    $\forall q > 0$, $\sup_{|\t| \le q}   \E(\abs{U(\t,Z)}^2) < \infty$.
  \end{hypo}
  Then, the sequence ${(\t_n)}_n$ defined by~(\ref{eq:chen}) converges a.s. to
  $\t^{\star}$ for any sequence of compact sets
  satisfying~(\ref{eq:chen-compact-def}) and moreover the sequence
  ${(\alpha _n)}_n$ is a.s. finite.
\end{theorem}

Note that the assumptions required to ensure the convergence are very weak and
are formulated independently of the algorithm trajectories, which makes them
easy to check. Since the variance reduction technique we settle here aims at
being automatic in the sense that it does not require any fiddling with the
gain sequence depending on the function $U$, it is quite natural to average
the procedure defined by Equation~(\ref{eq:chen}).

\subsection{Averaging a stochastic algorithm}
\label{sec:average}

This section is based on the remark that Cesaro type averages tend to smooth
the behaviour of convergent estimators at least from a theoretical point of
view. Such averaging techniques have already been studied and proved to
provide asymptotically efficient estimators (see for instance
\cite{polyak92:_accel}, \cite{kushner:1045} or \cite{MR1780908}).

At the same time, it is well known that true Cesaro averages are not so
efficient from a practical point of view because the rate at which the impact
of the first iterates vanishes in the average is too slow and it induces some
kind of a numerical bias which in turn dramatically slows down the
convergence. Combining these two facts has led us to consider a moving window
average of Algorithm~(\ref{eq:chen}).
\\
In this section, we restrict to gain sequences of the form $\gamma_n =
\frac{\gamma}{(n+1)^a}$ with $\frac{1}{2} < a < 1$. Let $\tau>0$ be
the length of the window used for averaging. For $n \ge 1$, we introduce
\begin{equation}
  \label{chen-average}
  \hat{\t}_n (\tau) = \frac{\gamma_{p}}{\tau} \sum_{i=p}^{p+\lfloor
    \tau/\gamma_{p} \rfloor} \t_i \quad \text{ with } p=\sup\{k \ge 1: k + \tau/\gamma_k
  \le n\} \wedge n.
\end{equation}
We use the convention $\sup \emptyset = +\infty$.    
The almost sure convergence of $(\hat{\t}_n (\tau))_n$ can easily be deduced
from Theorem~\ref{thm-chen}. The asymptotic normality of the sequence $(
\hat{\t}_n (\tau))_n$ has been studied in \cite{lelong_phd}. The definition of
$\hat\t_n$ is a little different from the one used in \cite{lelong_phd}
because we want to ensure that the sequence $(\hat\t_n)_n$ is adapted to the
filtration $\cf_n$ in view of the use of $\hat\t_n$ as an estimator of $\ts$ in
Algorithm~\ref{algo:coupled}.

\section{Examples of parametric Monte-Carlo settings}
\label{sec-ex}

In this section, we give various examples of cases in which a parametric
representation of the expectation of interest is available
\[
\E(Z)=\E(H(\t,X)).
\] In each example, we highlight the strong convexity and the regularity of
the function $\t \longmapsto \E(H^2(\t,X))$ such that the minimiser $\t^\s$ is
uniquely defined as the one root of $\t \longmapsto \nabla_\t \E(H^2(\t,X))$.



\subsection{Importance sampling for normal random variables}

Let $G=(G_1,\ldots, G_d)$ be a $d-$dimensional standard normal random
vector. For any measurable function $h : \R^d \longrightarrow \R$ such that
$\E(\abs{h(G)}) < \infty$, one has for all $\t\in\R^d$
\begin{equation}
  \label{girsanov}
  \E\left(h(G)\right) = 
  \E\left(e^{-\t \cdot G - \frac{|\t|^2}{2}}h(G+\t)\right).
\end{equation}
Assume we want to compute $\E(f(G))$ for a measurable function $f : \R^d
\longrightarrow \R$ such that $f(G)$ is integrable. By applying
equality~\eqref{girsanov} to $h=f$ and $h(x) = f^2(x) \expp{- \t \cdot x +
  \frac{\abs{\t}^2}{2}}$, one obtains that the expectation and the variance of the
random variable $f(G + \t) \expp{- \t \cdot G - \frac{\abs{\t}^2}{2}}$ are
respectively equal to $\E(f(G))$ and $v(\t) - \E^2(f(G))$ where
\begin{equation*}
  v(\t) = \E\left( f^2(G) \expp{-\t \cdot G + \frac{\abs{\t}^2}{2}}\right).
\end{equation*}
The strict convexity of the function $v$ is already known from
\cite{Su02optimalimportance} for instance. For the sake of completeness, we
prove here a slightly improved version of this result.
\begin{proposition} \label{convexity} Assume that
  \begin{align}
    \label{fnonul}
    &\P(f(G) \ne 0) > 0, \\
    \label{integ}
    &\exists \varepsilon>0, \; \E(\abs{f(G)}^{2+\varepsilon}) < \infty
  \end{align} Then, $v$ is infinitely continuously differentiable and strongly
  convex.
\end{proposition}
\begin{proof}
  The function $\t\mapsto f^2(G)e^{-\t \cdot G+\frac{|\t|^2}{2}}$
  is infinitely continuously differentiable.
  Since,
  \begin{align*}
    \sup_{|\theta|\leq M} |\partial_{\t^j}
    f^2(G)e^{-\t \cdot G+\frac{|\t|^2}{2}}|\leq
    e^{\frac{M^2}{2}}f^2(G)\left(M+(e^{G^j}+e^{-G^j})\right)
    \prod_{k=1}^d(e^{MG^k}+e^{-MG^k})
  \end{align*} where the right hand side is integrable because by Hölder's
  inequality and Equation~\eqref{integ}, we have $ \forall \t \in \R^d,
  \E\left( f^2(G) \expp{-\t \cdot G} \right) < \infty.$ Lebesgue's theorem
  ensures that $v$ is continuously differentiable with
  $\frac{\partial}{\partial_{\t^j}}v(\t)=\E\left(f^2(G)(\t^j-G^j)e^{-\t \cdot
      G+\frac{|\t|^2}{2}}\right)$.  Higher order differentiability properties
  are obtained by similar arguments and in particular the Hessian matrix
  writes
  \begin{align*}
    \nabla^2 v(\t) = \E\left(f^2(G) e^{-\t \cdot
    G+\frac{|\t|^2}{2}})\right) I  + \E\left( (\t-G)(\t-G)^* f^2(G) e^{-\t
    \cdot G+\frac{|\t|^2}{2}}\right)
  \end{align*}
  The second term in the above equation is a positive semi-definite matrix, hence 
  \begin{align*}
    \nabla^2 v(\t) \ge \E(f^2(G) e^{-\t \cdot G+\frac{|\t|^2}{2}}) 
    = \E(f^2(G) e^{-\t \cdot G}) \E(e^{\t \cdot G}) \ge \E(|f(G)|)^2.
  \end{align*}
  Assumption \eqref{fnonul} ensures that $\E(|f(G)|)>0$. Then, the Hessian
  matrix is uniformly bounded from below by the positive definite matrix
  $\E(|f(G)|)^2 I$. This yields the strong convexity of the function $v$.
\end{proof}
Proposition~\ref{convexity} implies that $v$ has a unique minimiser $\t^\s$
characterised by $\nabla v(\t^\s) = 0$, i.e.  $\E\left((\t^\s - G)
  e^{-\t^\s\cdot G +\frac{|\t^\s|^2}{2}} f^2(G) \right) = 0$.

\subsection{Importance sampling for processes}

Equality~\eqref{girsanov} can actually be extended to the Brownian motion
framework using Girsanov's theorem. Let $(W_t, 0 \le  t \le T)$ be a
$d-$dimensional Brownian motion and $\cf$ its natural filtration. For any
measurable and $\cf-$predictable process $(\t_t, 0 \le  t \le T)$ such that
$\E\left( \expp{ \frac{1}{2} \int_0^T \abs{\t_t}^2 dt} \right) < \infty$, one
has
\begin{equation*}
  \E\left(f(W_t,0\le t \le T)\right) = 
  \E\left(e^{-\int_0^T \t_t \cdot dW_t -
      \frac{1}{2} \int_0^T \abs{\t_t}^2 dt}
    f\left(W_t + \int_0^t \t_s ds ,0 \le t \le T\right)\right).
\end{equation*}
Assume $\t_t = \t \in \R^d$, for all $t \in [0, T]$. The
variance of $\expp{-\t \cdot W_T - \frac{\t^2 T}{2}} f(W_t,0\le t \le T)$
  writes down $v(\t) - \E\left(f^2(W_t,0\le t \le T)\right)$ with
\begin{equation*}
  v(\t) = \E\left(e^{-\t \cdot W_T + \frac{\abs{\t}^2}{2} T }
    f^2\left(W_t + \t t,\, 0 \le t \le T\right)\right).
\end{equation*}
A similar result to Proposition~\ref{convexity} holds; in particular  $v$ is
infinitely continuously differentiable, strictly convex and goes to infinity
at infinity. 

For more general processes $(\t_t, 0 \le t \le T)$, we refer the reader
to~\cite{lemairepages08}.

\subsection{The exponential change of measure}

The idea of tilting some probability measure to find the ones that minimises
the variance is a very common idea which can be also be applied to a wide
range of distribution, see for instance the recent results of Kawai
\cite{kawai09-1,kawai08-1} in which he applied an exponential change of
measure to Lévy processes, also known as the Esscher transform.

Consider a random variable $X$ with values in $\R^d$ and cumulative generating
function $\psi(\t) = \log \E\big(\expp{\t \cdot X}\big)$. We assume that
$\psi(\t) < \infty$ for all $\t \in \R^d$. Let $p$ denote the density of
$X$. We define the density $p_{\t}$ by
\begin{equation*}
  p_{\t} (x) = p(x) \expp{\t \cdot x - \psi (\t)}, \quad x \in \R^d.
\end{equation*}
Let $X^{(\t)}$ have $p_\t$ as a density, then
\begin{equation*} 
  \E(f(X))  = \E \left[f(X^{(\t)}) \frac{p(X^{(\t)})}{p_\t(X^{(\t)})}
  \right].
\end{equation*}
The variance of $f(X^{(\t)}) \frac{p(X^{(\t)})}{p_\t(X^{(\t)})}$ writes $v(\t)
- \E\Big(f(X^{(\t)})^2 \frac{p(X^{(\t)})^2}{p_\t(X^{(\t)})^2}\Big)$ with
\begin{equation*}
  v(\t) = \E \left(\left| f(X^{(\t)}) \frac{p(X^{(\t)})}{p_\t(X^{(\t)})} \right|^2
  \right) = \E \left( f(X)^2 \expp{- \t \cdot X + \psi(\t)} \right).
\end{equation*}
Obviously, this change of measure is only valuable as a variance reduction
technique if $X^{(\t)}$ can be simulated at approximately the same cost as $X$.
\begin{proposition}
   Assume that
  \begin{align}
    \label{integ-esscher}
    &\exists \varepsilon>0, \; \E(\abs{f(G)}^{2+\varepsilon}) < \infty \\
    \label{p-theta}
    & \lim_{|\t| \longrightarrow \infty} p_\t(x) = 0 \text{ for all $x$ in $\R^d$}
  \end{align} Then, $v$ is infinitely continuously differentiable, convex and
  $\lim_{|\t| \longrightarrow \infty} v(\t) = \infty$.
\end{proposition}
\begin{proof} To prove the differentiability of $v$, it suffices to reproduce
  the first part of the proof of Proposition~\ref{convexity}.
  The convexity of $v$ comes from the log-convexity of $\psi$. Moreover,
  \begin{equation*}
    v(\t) = \E \left( f(X)^2 \expp{- \t \cdot X + \psi(\t)} \right) = \E
    \left( f(X)^2 \frac{p(X)}{p_\t(X)} \right)
  \end{equation*}
  Combining Equation~\eqref{p-theta} with Fatou's Lemma yields that
  $\lim_{|\t| \longrightarrow \infty} v(\t) = \infty$.
\end{proof}

\begin{remark}
  If $X$ is a random standard normal vector, $p_{\t}(x) = p(x - \t)$ and
  $X^{(\t)}$ is a random normal vector with mean $\t$ and identity covariance
  matrix. Hence, we recover Equality~\eqref{girsanov}.
\end{remark}

\section{Application to the Gaussian random vector framework}

\subsection{Presentation of the problem}
\label{gaussianframework}

We consider a $D-$multidimensional local volatility model in which each asset
is supposed to be driven by the following dynamics under the risk neutral
measure.
\begin{equation*}
  dS^i_t = S^i_t( r dt + \sigma(t,S^i_t)\cdot dW^i_t), \ S^i_{0}=s^i.
\end{equation*}
$W=(W^1, \dots, W^{D})^*$ is a vector of correlated standard Brownian
motions. The covariance structure of the Brownian motions is given by $\langle
W, W \rangle_t = \Gamma t$ where $\Gamma$ is a definite positive matrix with a
diagonal filled with ones. In our numerical examples, we take $\Gamma_{ij} =
\ind{i=j} + \rho \ind{i \neq j}$ with $\rho \in (\frac{-1}{D-1}, 1)$ to ensure
that the matrix $\Gamma$ is positive definite. The function $\sigma$ is the
local volatility function, $r$ is the instantaneous interest rate and the
vector $(s^1, \dots, s^{D})$ is the vector of the spot values. In this model,
we want to price path-dependent options whose payoffs can be written as a
function of $(S_t,t\le T)$. Hence, the price is given by the discounted
expectation $\expp{-r T} \E(\psi(S_t,t\le T))$. Most of the time, this
expectation must be computed by Monte Carlo methods and one has to consider an
approximation of $\psi(S_t,t\le T)$ on a time grid $0=t_0 < t_1 < \dots < t_N
= T$. Then, the quantity of interest becomes
\begin{equation*}
  \expp{-rT} \E(\hat{\psi}(S_{t_0}, S_{t_1}, \ldots, S_{t_N})).
\end{equation*}
The discretisation of the asset $S$ can for instance be obtained using an
Euler scheme, which means that the function $\hat{\psi}$ can be expressed in
terms of the Brownian increments or equivalently using a random normal
vector. These remarks finally turn the original pricing problem into the
computation of an expectation of the form  $\E(\phi(G))$ where $G$ is a
standard normal random vector in $\R^{N D}$ and $\phi : \R^{N \times D}
\longmapsto \R$ is a measurable and integrable function. Using
Equation~\eqref{girsanov}, we have for all $\t \in \R^d$,
\begin{equation}
  \label{Agirs}
  \E(\phi(G)) = \E\left(\phi(G + A \t ) e^{- A \t \cdot G -
      \frac{|A \t|^2}{2}} \right),
\end{equation}
where $A$ is $d \times ND$ matrix. The particular choice $d = ND$ and $A =
I_d$ corresponds to Equation~\eqref{girsanov}. When $D=1$, the choice $d=1$
and $A = (\sqrt{t_1}, \sqrt{t_2 - t_1}, \dots, \sqrt{t_N-t_{N-1}})^*$
corresponds to adding a linear drift to the one dimensional standard Brownian
motion $W$ and we recover the Cameron-Martin formula.

Transformation~\eqref{Agirs} actually relies on an importance sampling change of
measure. Other strategies may be applicable such as stratification for
instance as it is explained by Glasserman et~al. in~\cite{glas_strati}. 

\subsection{Bespoke estimators for the optimal variance parameter}

It ensues from Proposition~\ref{convexity}, that the second moment
\begin{equation}
  \label{vA}
  v(\t)  = \E\left(\psi(G + A \t )^2 e^{-2 A \t
      \cdot G - \abs{A \t}^2} \right) = \E\left( \phi(G)^2 e^{- A \t \cdot G +
      \frac{\abs{A \t}^2}{2}} \right)
\end{equation}
is strongly convex, infinitely differentiable and 
\begin{equation}
  \label{nablav1}
  \nabla v(\t) = \E\left( A^*(A \t - G) \phi(G)^2 e^{- A \t \cdot G +
      \frac{\abs{A \t}^2}{2}} \right).
\end{equation}
If we apply Equation~\eqref{girsanov} again, we obtain an other expression for
\begin{equation}
  \label{nablav2}
  \nabla v(\t) = \E\left( -A^* G \phi(G + A \t)^2 e^{- 2 A \t \cdot G +
      \abs{A \t}^2} \right).
\end{equation}
Let us introduce the following two functions
\begin{align}
  \label{U1}
  U^1(\t,G) &= A^*(A \t - G) \phi(G)^2 e^{- A \t \cdot G +
    \frac{\abs{A \t}^2}{2}}, \\
  \label{U2}
  U^2(\t, G) &= -A^* G \phi(G + A \t)^2 e^{- 2 A \t \cdot G +
      \abs{A \t}^2}.
\end{align}
Using either Equation~\eqref{nablav1} or~\eqref{nablav2}, we can write $\nabla
v(\t) = \E(U^2(\t,G)) = \E(U^1(\t,G))$ and these two functions $U^1$ and $U^2$
fit in the framework of Section~\ref{sec-sa} and enable to construct two
estimators of $\ts$ $(\t^1_n)_n$ and $(\t^2_n)_n$ following
Equation~\eqref{eq:chen}
\begin{align}
  \label{theta1}
  \t^1_{n + 1} & = {\cal T}_{K_{\alpha_n}} \left(\t^1_{n} - \gamma_{n+1}
    U^1(\t^1_n, G_{n+1})\right),\\
  \label{theta2}
  \t^2_{n + 1} & = {\cal T}_{K_{\alpha_n}} \left(\t^1_{n} - \gamma_{n+1}
    U^2(\t^2_n, G_{n+1})\right),
\end{align}
where $G_n$ is an i.i.d sequence of random variables following the law of $G$.
We also introduce their corresponding averaging versions
$(\widehat{\t^1}_n)_n$ and $(\widehat{\t^2}_n)_n$ following
Equation~\eqref{chen-average}.  Based on Equation~\eqref{Agirs}, we define
\begin{equation*}
  H(\t, G) = \phi(G + A \t ) e^{- A \t \cdot G - \frac{|A \t|^2}{2}}.
\end{equation*}
Corresponding to the different estimators of $\ts$ listed above, we can define
as many approximations of $\E(\phi(G))$ following
Equation~\eqref{adaptive-estim}
\begin{align*}
  \xi^1_n = \frac{1}{n} \sum_{i=1}^n H(\t^1_{i-1}, G_i), \quad
  &\xi^2_n = \frac{1}{n} \sum_{i=1}^n H(\t^2_{i-1}, G_i),\\
  \hat\xi^1_n = \frac{1}{n} \sum_{i=1}^n H(\hat\t^1_{i-1}, G_i), \quad
  & \hat\xi^2_n = \frac{1}{n} \sum_{i=1}^n H(\hat\t^2_{i-1}, G_i),
\end{align*}
where the sequence $G_i$ has already been used to build the $(\t_n)_n$
estimators. From Proposition~\ref{convexity} and Theorems~\ref{thm-chen},
\ref{adapslln} and \ref{adaptcl}, we can deduce the following result.
\begin{theorem}
  \label{thm:is-chen} If there exists $\varepsilon>0$ such that
  $\E(\phi(G)^{4+\varepsilon}) < \infty$ then, the sequences $(\t^1_n)_n$,
  $(\t^2_n)_n$, $(\widehat{\t^1}_n)_n$ and $(\widehat{\t^2}_n)_n$ defined by
  Equations~\eqref{eq:chen} or~\eqref{chen-average} converge a.s. to
  $\t^\star$ for any increasing sequence of compact sets $(K_j)_j$
  satisfying~(\ref{eq:chen-compact-def}) and the adaptive estimator $(\xi^1_n)_n,
  (\xi^2_n)_n,(\hat\xi^1_n)_n,(\hat\xi^2_n)_n$ converge to $\E(\phi(G))$ and are
  asymptotically normal with optimal limiting variance $v(\t^\s)$.
\end{theorem}
\begin{proof}
  We only do the proof for $(\t^1_n)_n$ and $(\widehat{\t^1}_n)_n$ as the same
  ideas can be applied to $(\t^2_n)_n$ and $(\widehat{\t^2}_n)_n$.
  We know from Proposition~\ref{convexity}, that the function $v$ defined by
  Equation~\eqref{vA} is strongly convex and infinitely differentiable, hence
  $\nabla v$ satisfies Assumption~\hypref{unique-rm}.
  Let $q > 0$. For any $\t$ satisfying $|\t| \le q$, we have
  \begin{align*}
    \E\left| U^1(\t,G) \right|^2 & \le \E \left( (\norm{A} (\norm{A} q + 
      |G|)^2 \phi(G)^4 e^{-2 A \t \cdot G}
      e^{\norm{A}^2 q^2} \right).
  \end{align*}
  Using Hölder's inequality, it can easily be proved that the expectation on
  the right hand side is uniformly bounded for $|\t| \le q$. Hence
  Assumption~\hypref{localbound} is satisfied. Therefore, $\t^1_n$ and $\hat
  \t^1_n$ both converge to $\ts$. Let $\eta >0$,
  \begin{align*}
    \E\left| H(\t,G) \right|^{2+\eta} & = \E \left( |\phi(G)|^{2+\eta} e^{-
        (1+\eta) A \t \cdot G}  e^{\frac{|A\t|^2(1+\eta)}{2}} \right).
  \end{align*}
  Using the integrability of $\phi(G)$ and Hölder's inequality, one can prove
  that the expectation on the .r.h.s is bounded for $\t$ in a ball. Moreover,
  combining this with Lebesgue's theorem, we obtain that the functions $\t
  \longmapsto \E|H(\t,G)|^2$ and $\t \longmapsto \E|H(\t,G)|^{2+\eta}$ are
  continuous. Therefore, the convergence and asymptotic normality of $\xi_n$
  issues from Theorems~\eqref{adapslln} and~\eqref{adaptcl}.
\end{proof}

\begin{remark} \label{rem-arouna} Theorem~\ref{thm:is-chen} extends the result
  of \cite[Theorem 4]{arouna03:_robbin_monro}. Our result is valid for any
  increasing sequences of compact sets $(K_j)_j$
  satisfying~(\ref{eq:chen-compact-def}) whereas Arouna needed a condition on
  the compact sets to ensure the convergence of the $(\t_n)_n$ estimators. The
  only condition required is some integrability on the payoff function and
  nothing has to be checked along the algorithm paths, which is a great
  improvement from a practical point of view.
\end{remark}
For the vast majority of payoff functions commonly used, the assumptions of
Theorem~\ref{thm:is-chen} are always satisfied.

\subsection{Numerical results}

\subsubsection{Complexity of the different approximations}
\label{cputimes}

In the introduction, we have presented two different strategies for implementing a
variance reduction method based on the approximation of the optimal variance
parameter. We know from Theorem~\ref{thm:is-chen}, that the adaptive and non-adaptive
algorithms both converges at the same rate and the same limiting variance. Therefore,
to decide which one is the better, we have to compare their computational costs. In
this section, we assume that the computational cost of the different algorithms is
determined by the number of evaluations of the function $\phi$. We will see in the
examples later that this assumption is realistic and therefore it becomes obvious
that the averaging and non-averaging estimators of $\ts$ all have the same
computational costs when implemented with expertise.

\paragraph{The non-adaptive algorithm}

We know from~\cite{arouna03:_robbin_monro,Su02optimalimportance} that the sequential
algorithm converges with a rate of $\sqrt{v(\ts)/n}$ if we have $2n$ samples at hand
and want to implement the sequential algorithm by using the first $n$ samples for
approximating $\ts$ and the last $n$ samples for actually computing the Monte Carlo
estimator with the previously computed approximation of $\ts$. Whatever approximation
of $\ts$ is used, be it $(\t^1_n)_n$, $(\t^2_n)_n$, $(\widehat{\t^1}_n)_n$ or
$(\widehat{\t^2}_n)_n$, this algorithm requires $2n$ evaluations of the function
$\phi$ whereas a crude Monte Carlo method only evaluates the function $\phi$ $n$
times and achieves a convergence rate of $\sqrt{v(0)/n}$, hence this method only
becomes efficient when $v(\ts) \le v(0)/2$.

\paragraph{The adaptive algorithm}

From Theorem~\ref{thm:is-chen}, we know that the adaptive estimators
$(\xi^1_n)_n, (\xi^2_n)_n,(\hat\xi^1_n)_n,(\hat\xi^2_n)_n$ all converge with
the same rate $\sqrt{v(\ts)/n}$ but as we will see it they do not have the
same computational cost. First, let us concentrate on $(\xi^1_n)_n$ and
$(\hat\xi^1_n)_n$, at each iteration $i$, the function $\phi$ has to be
computed twice~: once at the point $G_{i+1}+\t^1_i$ (or $G_{i+1}+\hat\t^1_i$)
to update the Monte Carlo estimator and once at the point $G_{i+1}$ to update
$\t^1_{i+1}$ or $\hat\t^1_{i+1}$. Hence, the computation of $\xi^1_n$ or
$\hat\xi^1_n$ requires $2n$ evaluations of the function $\phi$. Similarly, the
computation of $\hat \xi^2_n$ requires at each step $2$ evaluations of the
function $\phi$~: one at the point $G_{i+1} + \hat\t^2_i$ to update the Monte
Carlo estimator and one at the point $G_{i+1} + \t^2_i$ to update the
stochastic algorithm. So the overall cost is still $2n$ evaluations of the
function $\phi$. But looking closely at the computation of $(\xi^2_n)_n$
immediately highlights the benefit of having put the parameter $\t$ back into
the function $\phi$ in the expression of $\nabla v$ : the updates of
$\xi^2_{i+1}$ and $\t^2_{i+1}$ both use the evaluation of the function $\phi$
at the same point $G_{i+1} + \t^2_i$. Hence, the computation of $\xi^2_n$ only
needs $n$ evaluations of the function $\phi$ instead of $2n$ for all the
others algorithms. Obviously, the computational costs of the different
estimators cannot really be reduced to the number of times the function $\phi$
is evaluated so one should not expect that computing $\xi^2_n$ is twice less
costly than the other estimators but we will see in the examples below that
the estimator $\xi^2_n$ is indeed faster than the others.
\\

To shortly conclude on the complexity of the different algorithms, be they
sequential or adaptive, one should bear in mind that all the estimators except
$(\xi^2_n)_n$ roughly require twice the computational time of the crude
Monte-Carlo method.

\subsubsection{Practical implementation}

The choice of using Equations~\eqref{eq:chen} or~\eqref{chen-average} to build an
estimator of $\ts$ becomes really important when one has to implement the variance
reduction procedure either by using Algorithms~\eqref{algo:not-coupled}
or~\eqref{algo:coupled}. Both the averaging and non-averaging strategies have pros
and cons. The averaging algorithm theoretically converges a little slower but has a
much smoother behaviour with respect to the proper adjustment of the gain sequence
$(\gamma_n)_n$. Then, to have a robust estimator \textemdash\ in the sense that the
numerical convergence of the estimator does not depend too much on the choice of of
the gain sequence \textemdash\ the averaging procedure proves to be better in
practice. The non averaging algorithm should converge a little faster even though we
do not notice it in practise as the convergence oscillates too much and is far more
sensitive to the proper choice of the sequence $(\gamma_n)_n$. Eventually, both
algorithms produce very similar results regarding variance reduction; the averaging
one is easier to tune but requires more computational time.  \\

In the numerical experiments of this section, we compare the different algorithms on
multi-asset options.  The quantity ``Var MC'' denotes the variance of the crude Monte
Carlo estimator computed on-line on a single run of the algorithm. The variance
denoted ``Var $\xi^2$'' (resp. ``Var $\hat \xi^2$'') is the variance of the ADIS
algorithm (see Algorithm~\ref{algo:coupled}) which uses $(\t^2_n)_n$ (resp.
$(\hat\t^2_n)_n$) to estimate $\ts$. These variances are computed using the on-line
estimator given by Corollary~\ref{varestim}. These adaptive algorithms are also
compared to the sequential strategy described by Algorithm~\ref{algo:not-coupled}
denoted by ``$\t^2$+MC'' or ``$\hat\t^2$+MC'' depending on how $\ts$ is approximated.
In all these algorithms, the matrix $A$ introduced in Equation~\eqref{Agirs} is
chosen as the identity matrix. When $A$ is not the identity matrix, its purpose is to
reduce the dimension of the space in which the optimal $\ts$ is searched and in such
cases the algorithms will be call ``reduced''.  Note that for the comparison to be
fair between the different strategies, we have used $n$ samples for the adaptive
algorithms but $2n$ for the sequential algorithms so that they all satisfy a central
limit theorem with the rate $\sqrt{n}$.

\paragraph{Basket options}

We consider options with payoffs of the form $(\sum_{i=1}^d \omega^i S_T^i - K)_+$
where $(\omega^1, \dots, \omega^d)$ is a vector of algebraic weights (enabling us to
consider exchange options). 
\begin{table}[h!t]  
  \centering\begin{tabular}{c@{$\quad$}c@{$\quad$}c@{$\quad$}c
      @{$\quad$}c@{$\quad$}c@{$\quad$}c}
    \hline\\
    $\rho$ & $K$ & $\gamma$ & Price & Var MC & Var $\xi^2$ & Var $\hat\xi^2$ \\
    \hline
    0.1 & 45 & 1   & 7.21 & 12.24 & 1.59  & 1.10 \\
        & 55 & 10  & 0.56 & 1.83  & 0.19  & 0.14 \\
    0.2 & 50 & 0.1 & 3.29 & 13.53 & 1.82  & 1.76 \\
    0.5 & 45 & 0.1 & 7.65 & 43.25 & 6.25  & 4.97 \\
        & 55 & 0.1 & 1.90 & 14.74 & 1.91  & 1.4  \\
    0.9 & 45 & 0.1 & 8.24 & 69.47 & 10.20 & 7.78 \\
        & 55 & 0.1 & 2.82 & 30.87 & 2.7   & 2.6  \\
    \hline    
  \end{tabular}
  \caption{Basket option in dimension $d=40$ with
    $r=0.05$, $T=1$, $S_0^i = 50$, $\sigma^i = 0.2$, $\omega^i = \frac{1}{d}$
    for all $i=1,\dots,d$ and $n=100\,000$.}
  \label{tab-basket}
\end{table}

\begin{table}[h!t]  
  \centering\begin{tabular}{c@{$\quad$}c@{$\quad$}c@{$\quad$}c@{$\quad$}c}
    \hline\\
    Estimators & MC   & $\xi^2$ & $\hat\xi^2$ \\
    \hline
    CPU time   & 0.85 & 0.9       & 1.64         \\
    \hline
  \end{tabular}
  \caption{CPU times for the option of Table~\ref{tab-basket}.}
  \label{tab-cpu-basket}
\end{table}

The results of Table~\ref{tab-basket} indicate that the adaptive algorithm
using an averaging stochastic approximation outperforms not only the crude
Monte Carlo approach but also the adapted algorithms using non-averaging
stochastic approximation. The better performance of the algorithms using
averaging estimators of $\ts$ comes from the better smoothness of the
averaging algorithm (see Equation~\eqref{chen-average}). Nonetheless, these
good results in terms of variance reduction must be considered together with
their computation costs reported in Table~\ref{tab-cpu-basket}. As explained
in Section~\ref{cputimes}, we notice that the computational cost of the
estimator $\xi^2$ is very close to the one of the crude Monte Carlo estimator
because the implementation made the most of the fact that the updates of
$\xi^2_{i+1}$ and $\t^2_{i+1}$ both need to evaluate the function $\phi$ at
the same point. Note that, because this implementation trick cannot be applied
to $\hat \xi^2$, the adaptive algorithm using an averaging stochastic
approximation is twice slower. For a given precision, the adaptive
algorithm is between $5$ and $10$ times faster.

\paragraph{Barrier Basket Options}

We consider basket options in dimension $D$ with a discrete barrier on each
asset. For instance, if we consider a Down and Out Call option, the payoff
writes down $(\sum_{i=1}^D \omega^i S_T^i - K)_+ \ind{\forall i \le D, \;
  \forall j \le N, \; S_{t_j}^i \ge L^i}$ where $\omega = (\omega^1, \dots,
\omega^D)$ is a vector of positive weights, $L=(L^1, \dots, L^D)$ is the
vector of barriers, $K>0$ the strike value and $t_N=T$. We consider one time
step per month, which means that for an option with maturity time $T=2$, the
number of time steps is $N=24$. From now on, we fix $D=5$. Hence if we use the
identity matrix $A$, the parameter $\t$ is of size $N \times D = 120$. Here,
we propose to reduce the dimension of $\t$ and we will in
Table~\ref{tab-basket-barrier5} that it achieves almost the same variance
reduction. Of course the matrix $A$ cannot be chosen independently of the
structure of the problem. Remember that the vector $G$ actually corresponds to
the increments of the Brownian motion $B$ with values in $\R^d$ on the grid
$(t_k = k T/N, k=0,\dots,N)$. We recall that we can simulate the Brownian
motion $B$ on the time grid $(t_k)_k$ by using the following equality in law
$$
\begin{pmatrix}
  B_{t_1} \\ B_{t_2} \\ \vdots \\ B_{t_{N-1}} \\ B_{t_N}
\end{pmatrix}
= \begin{pmatrix}
  \sqrt{t_1} I_D & 0 & 0 &\hdots &0\\
  \sqrt{t_1} I_D &\sqrt{t_2-t_1}  I_D& 0 &\hdots &0\\
  \vdots&\ddots&\ddots&\ddots&\vdots\\
  \vdots&\ddots&\ddots& \sqrt{t_{N-1}-t_{N-2}} I_D &   0 \\
  \sqrt{t_1} & \sqrt{t_2-t_1} I_D &\hdots & \sqrt{t_{N-1}-t_{N-2}} I_D
  &\sqrt{t_N-t_{N-1}} I_D
\end{pmatrix}
G
$$
where $I_D$ is the identity matrix in dimension $D$.  If we choose
$$
A = \begin{pmatrix}
  \sqrt{t_1} I_D \\
  \sqrt{t_2-t_1} I_D \\
  \vdots \\
  \vdots \\
  \sqrt{t_N-t_{N-1}} I_D
\end{pmatrix}
$$
then the transformation $G + A \t$ corresponds to the transformation $(B_{t_1}
+ \t t_1, B_{t_2} + \t t_2, \dots, B_{t_N} + \t t_N)^*$ and it reduces the
effective dimension of the importance sampling parameter to $D=5$ rather than
$D N = 120$.

\begin{table}[h!t]  
  \centering\begin{tabular}{c@{$\quad$}c@{$\quad$}c@{$\quad$}c@{$\quad$}c
      @{$\quad$}c@{$\quad$}c@{$\quad$}c@{$\quad$}c@{$\quad$}c}
    \hline\\
    $K$ & $\gamma$ & Price & Var  MC &  Var $\xi^2$ &  Var $\hat\xi^2$ 
    & Var   & Var $\xi^2$ &  Var $\hat\xi^2$ & Var \\
    &  &  &  &  &  & $\hat\t^2$+MC &         &         & $\hat\t^2$+MC \\
    &  &  &  &  &  &               & reduced & reduced & reduced       \\
    \hline
    45 & 0.5 & 2.37 & 22.46 & 4.92 & 3.52 & 2.59 & 2.64 & 2.62 & 2.60 \\
    50 & 1   & 1.18 & 10.97 & 1.51 & 1.30 & 0.79 & 0.80 & 0.80 & 0.79 \\
    55 & 1   & 0.52 & 4.85  & 0.39 & 0.38 & 0.19 & 0.24 & 0.23 & 0.19 \\
    \hline
  \end{tabular}
  \caption{Down and Out Call option in dimension $I=5$ with
    $\sigma = 0.2$, $S_0=(50, 40, 60, 30, 20)$, $L=(40, 30, 45, 20 ,10)$,
    $\rho=0.3$, $r=0.05$, $T=2$, $\omega = (0.2,0.2,0.2,0.2,0.2)$
    and $n=100\,000$.}
  \label{tab-basket-barrier5}
\end{table}
\begin{table}[h!t]  
  \centering\begin{tabular}{c@{$\quad$}c@{$\quad$}c@{$\quad$}c@{$\quad$}
      c@{$\quad$}c@{$\quad$}c@{$\quad$}c@{$\quad$}c}
    \hline\\
    Estimators & MC   & $\xi^2$ & $\hat\xi^2$ & $\t^2$ + MC &
    $\xi^2$ & $\hat\xi^2$ & $\t^2$ + MC \\
    & &  & &  &  reduced & reduced & reduced \\
    \hline
    CPU time & 1.86 & 1.93 & 3.34 & 4.06 &  1.89 & 2.89 & 3.90\\
    \hline
  \end{tabular}
  \caption{CPU times for the option of Table~\ref{tab-basket-barrier5}.}
  \label{tab-cpu-barrierbasket}
\end{table}

First, we note from Table~\ref{tab-basket-barrier5} that the reduced and non-reduced
ADIS algorithm achieve almost the same variance reduction. Actually, it is even
advisable to reduce the size of the importance sampling parameter to reduce the noise
in the stochastic approximation and therefore in the adaptive Monte Carlo estimator.
Comparing the columns ``Var MC'', ``Var $\xi^2$'' and ``Var $\hat\xi^2$'' points out
that when the convergence of the estimator of $\ts$ is too slow the first iterates of
the adaptive Monte Carlo estimators use wrong values of $\t$ and therefore cannot
reach $v(\ts)$ whereas if a sequential algorithm is used all the iterates of the
Monte Carlo estimator use the same and better approximation of $\t$. We can see in
Table~\ref{tab-basket-barrier5} that the variance of ``$\hat\xi^2$+MC'' is half the
one of $\xi^2$ or $\hat \xi^2$ but its CPU time is twice the one of $\xi^2$ as noted
in Table~\ref{tab-cpu-barrierbasket}.

The reduced algorithms are a little faster than the non-reduced ones but their real
advantage is to converge much more stably and to achieve the same variance as
``$\hat\xi^2$+MC'' but in far less computational time. As in the previous examples,
we still observe that the estimator ``$\xi^2$ reduced'' is faster than the others and
has a variance very close to the best method which is ``$\hat \t^2$+MC''.

\section{Conclusion}

In this work, we have explained how one could devise an adaptive variance reduction
method for computing an expectation with a free parameter. Different algorithms have
been studied both from a theoretical point of view and in practice. Although all the
adaptive algorithms satisfy the same central limit theorem, they may behave very
differently in practice, in particular adaptive algorithms using a non-averaging
stochastic approximation of the optimal variance parameter can be implemented in a
clever way which makes them as fast as a crude Monte Carlo approach. Nevertheless,
the numerical convergence of these stochastic is very sensitive to the tuning of
their gain sequence and one way to smooth this behaviour is to plug an averaging
procedure on top of the stochastic approximation but then the computational time
significantly increases and yet the dependency with respect to the gain sequence is
still a serious drawback.  To encounter the fine tuning of the algorithm, Jourdain
and Lelong~\cite{jourdainlelong08} have recently suggested to use deterministic
optimisation techniques coupled with sample approximation, but their technique can
not be implemented in an adaptive manner which increases its computational cost.

\bibliographystyle{plain}
\bibliography{adaptive}
\end{document}